\documentclass[journal]{IEEEtran}

\ifCLASSINFOpdf
\else
\fi

\usepackage{etex}
\usepackage{algorithmicx}
\usepackage{amsmath,amssymb,amsthm} 
\usepackage{bbding}
\usepackage[noadjust]{cite}
\usepackage{pdfpages}
\usepackage{array}
\usepackage{graphicx}
\usepackage{xcolor,colortbl}
\usepackage{epstopdf}
\usepackage{setspace}
\usepackage{ctable}
\usepackage{enumerate}
\usepackage{ragged2e}
\usepackage{amssymb}
\usepackage{multirow}
\usepackage{tabularx,tabulary}
\usepackage{nicefrac}		%
\usepackage[caption=false,font=footnotesize,labelformat=simple]{subfig} %
\usepackage[flushleft]{threeparttable}
\usepackage[bookmarks=false]{hyperref}
\usepackage{rotating}
\usepackage{tabularx,ragged2e,booktabs}
\newcolumntype{C}[1]{>{\Centering}m{#1}}

\usepackage{mathtools}
\usepackage{balance}

\usepackage{authblk}
\usepackage{verbatim}

\usepackage{xcolor}
\usepackage{tcolorbox}
\newtcbox{\highlight}[0]{boxsep=0pt,left=0pt,top=0pt,bottom=0pt,right=0pt,boxrule=0pt,arc=0pt,auto outer arc,colback=green,width=15cm}

\definecolor{dark-blue}{RGB}{0,70,127}

\usepackage{pifont}
\usepackage{framed}
\usepackage{soul}
\usepackage{booktabs}
\usepackage{multirow}

\usepackage[export]{adjustbox}

\begin{document}

\title{HACK3D: Crowdsourcing the Assessment of Cybersecurity in Digital Manufacturing}

\author{
  \IEEEauthorblockN{
  Michael Linares\IEEEauthorrefmark{1}\IEEEauthorrefmark{3}\thanks{\IEEEauthorrefmark{3} The first two authors have contributed to this work equally.},
  Nishant Aswani\IEEEauthorrefmark{1}\IEEEauthorrefmark{3},
  Gary Mac\IEEEauthorrefmark{1},
  Chenglu Jin\IEEEauthorrefmark{2}\thanks{\IEEEauthorrefmark{2} Corresponding author. Email: chenglu.jin@cwi.nl},
  Fei Chen\IEEEauthorrefmark{1},
  Nikhil~Gupta\IEEEauthorrefmark{1}, and~Ramesh~Karri\IEEEauthorrefmark{1}~\IEEEmembership{Fellow,~IEEE}
  }
  
  \IEEEauthorblockA{\IEEEauthorrefmark{1}New York University, Tandon School of Engineering, Brooklyn, NY 11201 USA}
  \IEEEauthorblockA{\IEEEauthorrefmark{2} CWI Amsterdam, The Netherlands}
  
}

\maketitle

\begin{abstract}
Digital manufacturing (DM) cyber-physical system is vulnerable to both cyber and physical attacks. HACK3D is a series of crowdsourcing red-team-blue-team events hosted by the NYU Center for Cybersecurity to assess the strength of the security methods embedded in designs using DM.  This study summarizes the lessons learned from the past three offerings of HACK3D, including ingenious ways in which skilled engineers can launch surprising attacks on DM designs not anticipated before. A key outcome is a taxonomy-guided creation of DM security benchmarks for use by the DM community. 
\end{abstract}

\section{Introduction}\label{s:intro}

Digital manufacturing (DM) security is gaining attention due to the involvement of trusted, partially trusted, and untrusted parties in the supply chain~\cite{zeltmann2016manufacturing}. %
A survey and taxonomy of threats and vulnerabilities have been developed~\cite{mahesh2020survey}. This survey of taxonomies shows that numerous attack vectors exist for the DM process chain but only a few specialized security schemes are available for this complex cyber-physical system (CPS). For example, DM attack vectors and impacts have been discussed from a cyber-physical perspective in \cite{sturm7cyber}.  %
Similarly, a stealthy DM tool path modification attack can go undetected \cite{wells2014cyber}. These attacks highlight the need for improved quality controls, cybersecurity education, and development of DM security assessment. A methodology for detecting attacks on an artifact's intrinsic behavior is presented in \cite{vincenttrojan}.

The DM process model has been studied and a new ``federated'' information systems architecture for DM has been developed in~\cite{kim2015streamlining}. This architecture  establishes requirements for end-to-end information sharing, quality control, and performance assurance. \cite{yampolskiy2014intellectual} investigate Intellectual Property (IP) protection for outsourced manufacturing, and study an alternative model that incorporates third party process tuning experts. The paper presents a risk assessment focused on IP protection and makes recommendation to minimize risks of this model. Furthermore,~\cite{mcnulty2012toward} surveys significance of DM for national security. 

A variety of cybersecurity methods have been developed that are specific to DM. These methods include hiding features in the design files and printing high-quality parts from these files by unauthorized users  difficult~\cite{chen2017security}. Embedding identification codes inside parts has been explored~\cite{chen2019embedding}. These codes are obfuscated by breaking them into hundreds of segments and hiding these segments in numerous layers in the part ~\cite{chen2019obfuscation}.  This study reports the outcomes of a series of crowdsourcing events focused on understanding the strengths and weaknesses of the security methods  developed for DM. We will describe the HACK3D\footnote{https://www.csaw.io/hack3d} designs and the attack methods developed by the participants. We will conclude with lessons learned from the crowdsourcing-based HACK3D events.

\vspace{2mm}
\noindent {\bf Paper roadmap.} In Section \ref{ss:chain}, we provide an overview of DM CPS, challenges in developing cybersecurity solutions for DM and a taxonomy of DM cyber threats. Section~\ref{s:hack3d} presents HACK3D, a crowdsourced red-team-blue-team event that NYU CCS has been organizing since 2018 to evaluate DM security methods. Concluding remarks are  in Section \ref{s:conclusion}.

\section{The DM Cyber-Physical System}\label{ss:chain}

\begin{figure*} [t]
 \centering
 \includegraphics[width=0.8\textwidth]{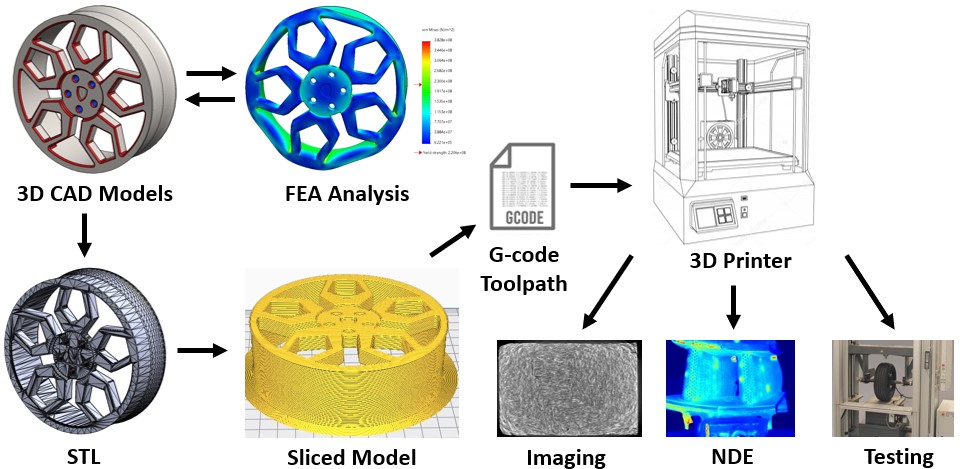}
 \caption{The digital manufacturing (DM) cyber-physical system makes use of connected systems such as 3D printers. STL refers to stereolithography file format and NDE refers to non-destructive evaluation methods such as radiography, ultrasonic imaging and tomography}
 \label{f:chain}
\end{figure*} 

\begin{figure}[t] 
 \centering
 \includegraphics[width = 3.5in]{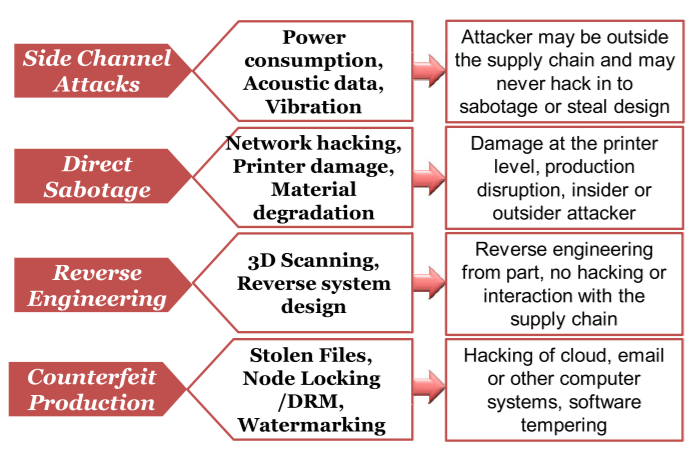}
 \caption{Classification of attacks in DM supply chain~\cite{gupta2020additive}.}
 \label{f:risk_class}
\end{figure}

\subsection{The Digital Manufacturing Process Chain}
Figure \ref{f:chain} shows the DM process chain that includes computer-aided design (CAD), design refinement by simulation tools such as finite element analysis (FEA), manufacturing of the part on a 3D printer followed by test and assembly. The product design process remains the same even in traditional manufacturing such as machining or milling. %
All steps involved in DM use computers and cloud for computation, collaboration, machine control, and data acquisition and analysis. Hence, all these steps are targets for cyber attacks.

\subsection{Taxonomy of cyber threats faced by DM}\label{s:taxonomy}

\begin{figure*}[t]
\centering
\includegraphics[width=\textwidth]{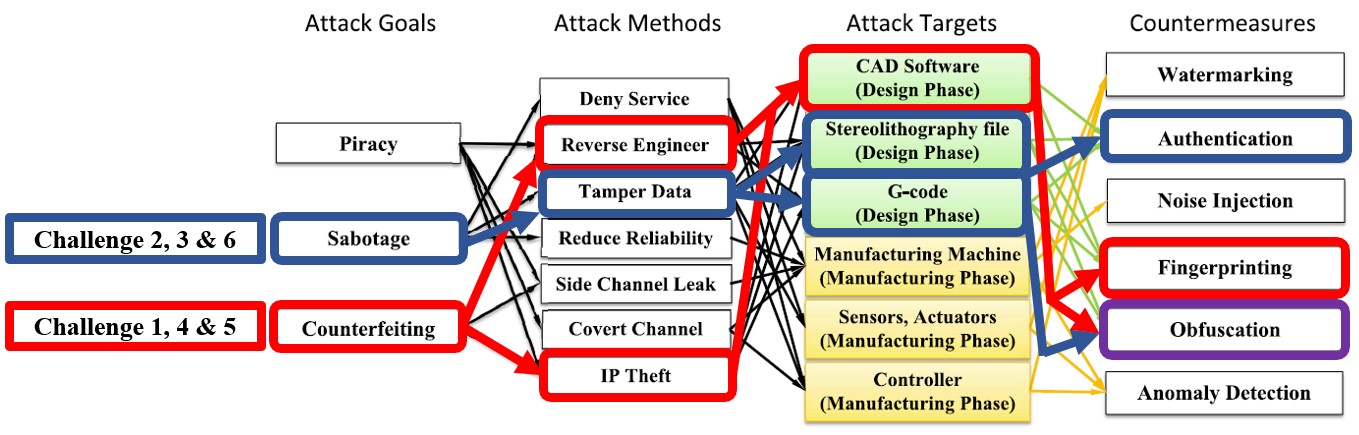}
\caption{The attack vectors that the HACK3D challenges demonstrated so far are highlighted in a taxonomy of security threats faced by DM from~\cite{mahesh2020survey}. }
\label{fig:attack_graph}
\end{figure*}

The attacks involved in the DM supply chain are classified in four categories as illustrated in Figure \ref{f:risk_class}  \cite{gupta2020additive}. For each category in the classification, different skills and tools are needed for a successful attack.  
As shown in Figure~\ref{fig:attack_graph}, the cybersecurity threats  present in the DM supply chain can be classified across four attack categories (goals, methods, targets, and countermeasures).
 Several security methods are available that can be applied to DM supply chain but strength of these methods for DM supply chain needs to be analyzed. %

\section{HACK3D Assesses Strength of DM Security}\label{s:hack3d}

An effective and widely used approach to assess the strength of security strategies is to conduct a red-team-blue team challenge involving participants from a diverse set of backgrounds. The objective of the HACK3D challenge is to provide a platform for these red-blue challenges to evaluate the robustness of new DM  security strategies. The HACK3D research team takes the role of the blue team, designing security methods for the manufacturing process, and presents them as HACK3D challenges. The approaches target a wide range of threats, including a focus on securing the digital design files. 

Red teams are crowd-sourced from a diverse population of students spanning all education levels and backgrounds. Their solutions provided a wide range of perspectives, some of which have not been considered when the blue team designed the security challenges. These security assessment benchmarks help determine the strengths and weaknesses of the blue team’s challenges, providing qualitative and quantitative insights for the design of future DM security policies and strategies. 

Each challenge investigates a pathway in the threat taxonomy from ~\cite{mahesh2020survey} as shown in Figure~\ref{fig:attack_graph}. In the preliminary rounds of HACK3D, the red teams had at least one month to solve the challenge. The final round was held onsite at New York University in 2018 and 2019 and became a virtual event in 2020 as part of the annual NYU Cybersecurity Awareness Week (CSAW) in early November. The participants had 1-2 days to solve the challenges in the final round.

\subsection{Challenge 1 (HACK3D 2018 Qualifying Challenge)}

\begin{figure*}[!t]
     \subfloat[\label{fig:hack3d_2018_target}]{%
       \includegraphics[width=0.5\columnwidth]{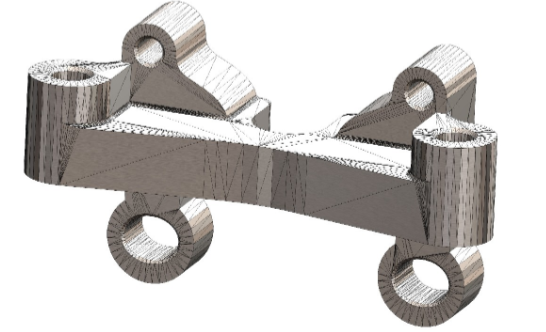}
     }
     \hfill
     \subfloat[\label{fig:hack3d_2018_point_cloud}]{%
       \includegraphics[width=0.5\columnwidth]{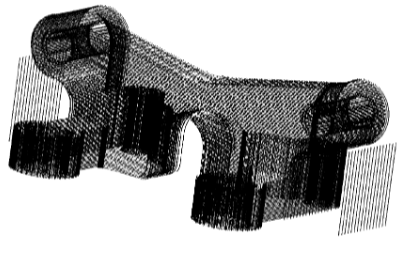}
     }
     \hfill
           \subfloat[\label{fig:hack3d_2018_final}]{%
      \includegraphics[width=0.9\columnwidth]{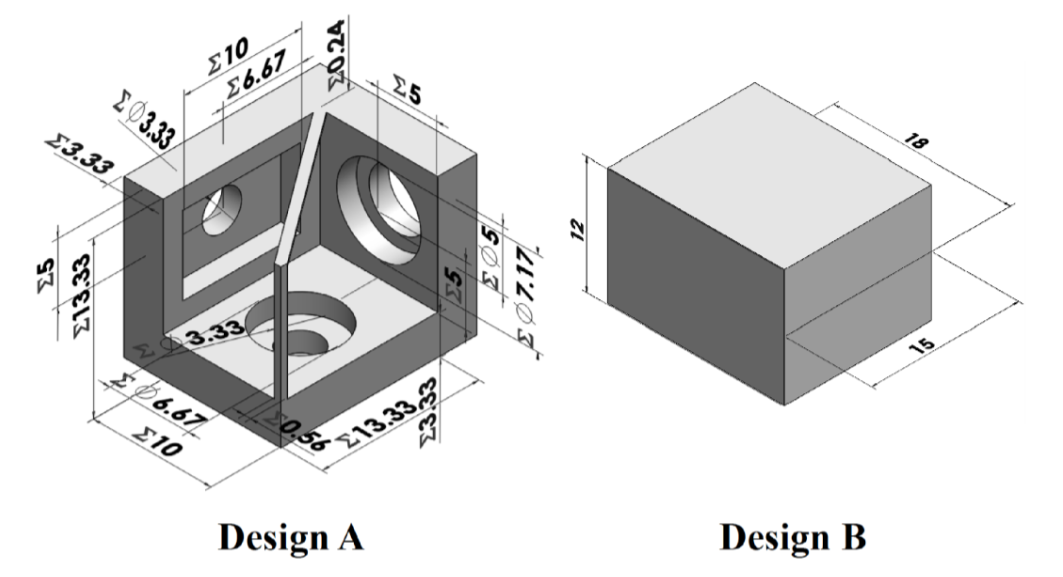}
     }
     \hfill
     \subfloat[\label{fig:hack3d_segments}]{%
      \includegraphics[width=0.45\columnwidth]{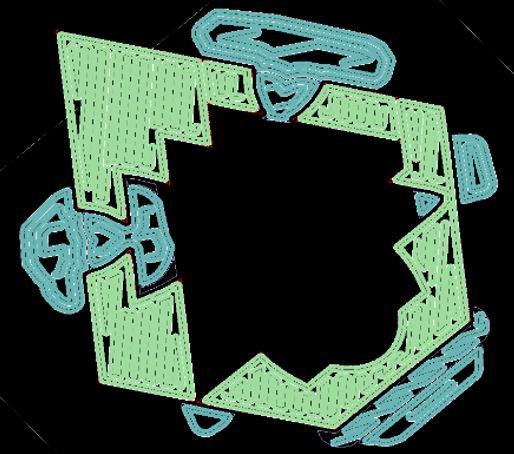}
     }
     \hfill
     \subfloat[\label{fig:hack3d_2018_spade}]{%
      \includegraphics[width=0.45\columnwidth]{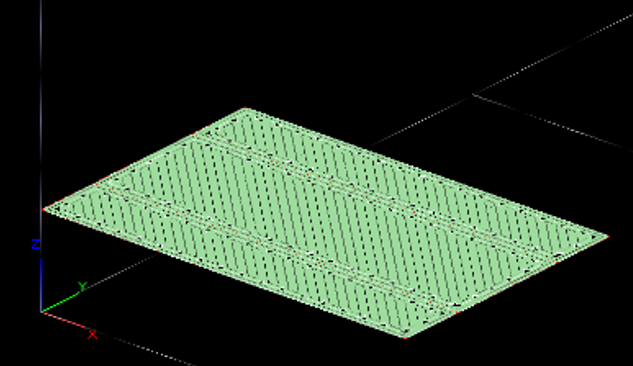}
     }
     \hfill
     \subfloat[\label{fig:hack3d_2018_finalA}]{%
      \includegraphics[width=0.45\columnwidth]{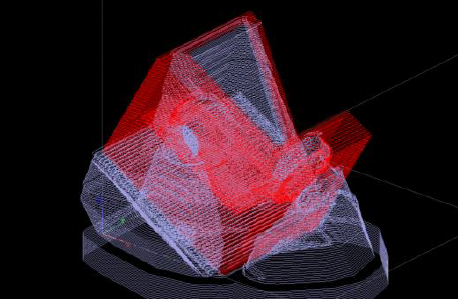}
     }
     \hfill
     \subfloat[\label{fig:hack3d_2018_finalB}]{%
      \includegraphics[width=0.45\columnwidth]{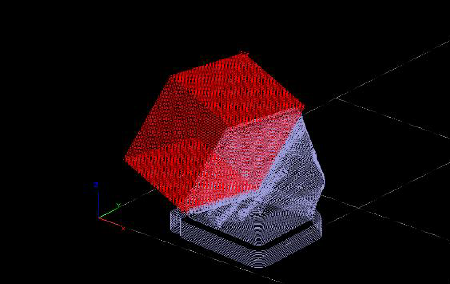}
     }
     \hfill
          \subfloat[\label{fig:chess_base}]{%
       \includegraphics[width=0.49\columnwidth]{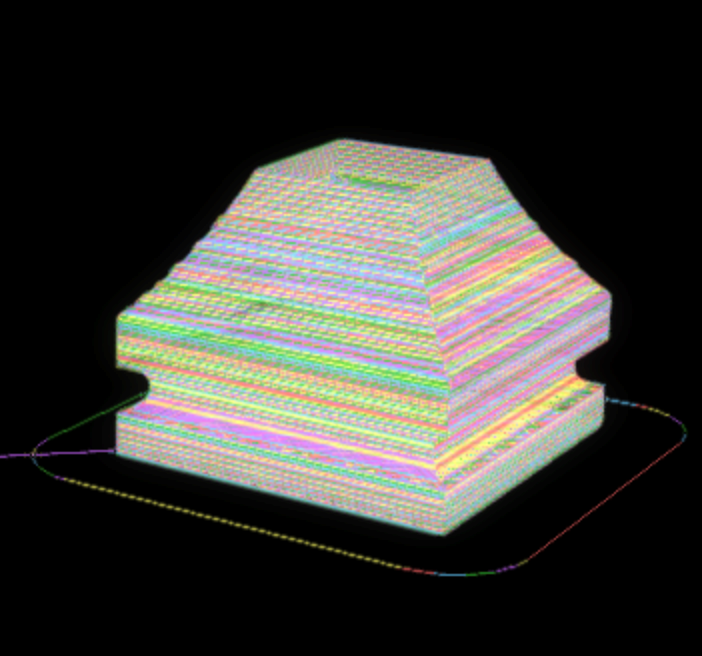}
     }
     \subfloat[\label{fig:chess_pieces}]{%
       \includegraphics[width=0.49\columnwidth]{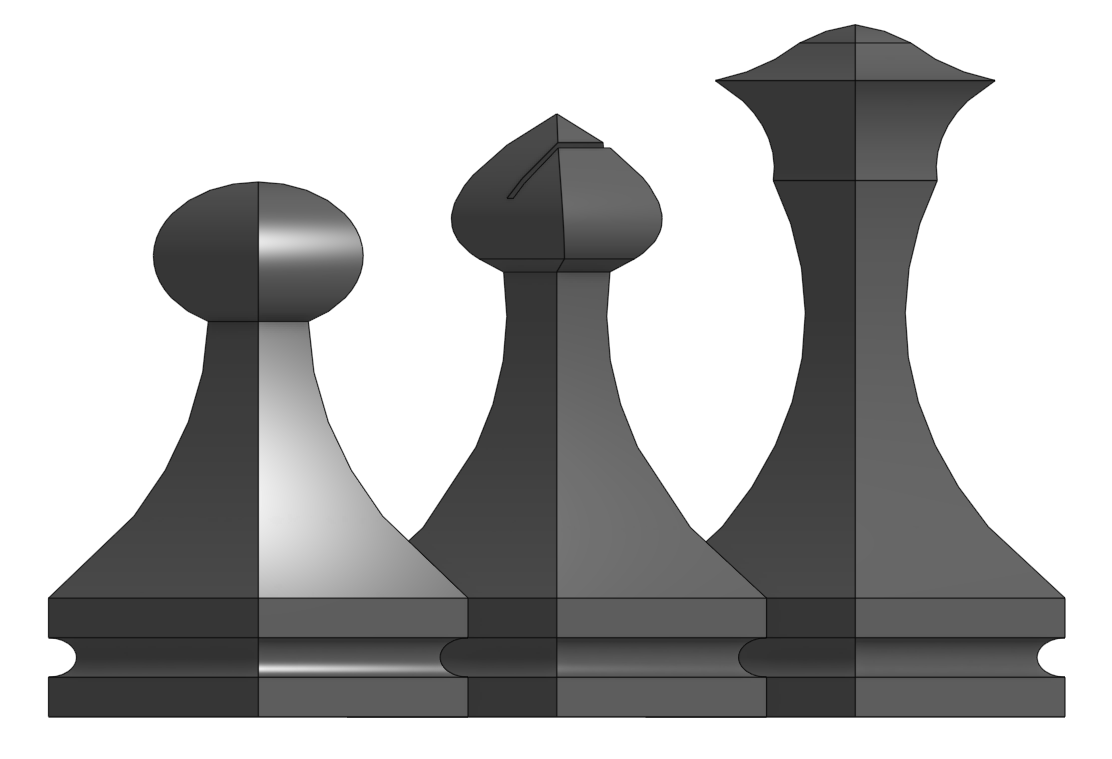}
     }
     \hfill
          \subfloat[\label{fig:spheres}]{%
       \includegraphics[width=0.49\columnwidth]{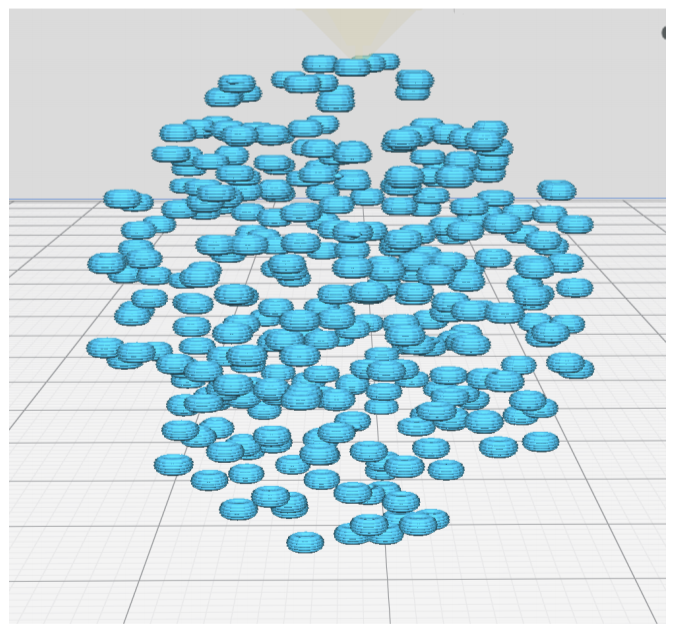}
     }
     \subfloat[\label{fig:qr_code}]{%
       \includegraphics[width=0.49\columnwidth]{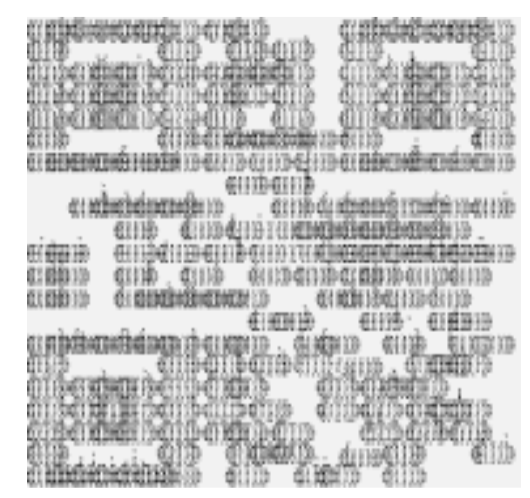}
     }
     \hfill
     \subfloat[\label{hack3d_challenge4}]{\includegraphics[width=\columnwidth]{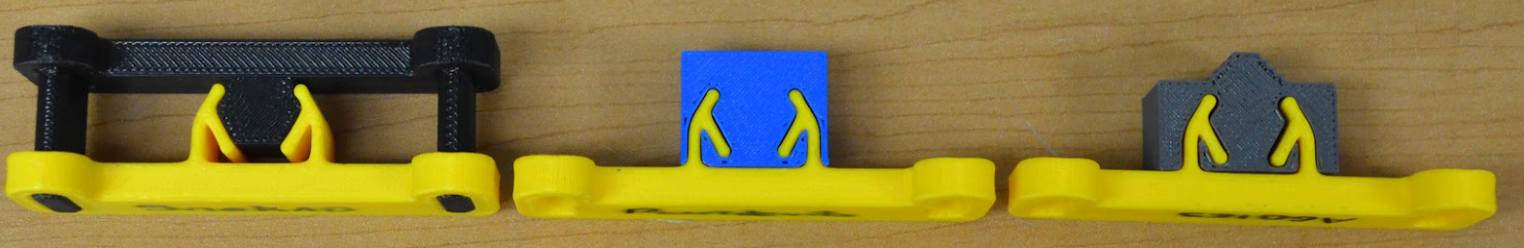}}
        \hfill
     \subfloat[\label{fig:hack3d2020_q2}]{%
       \includegraphics[width=\columnwidth]{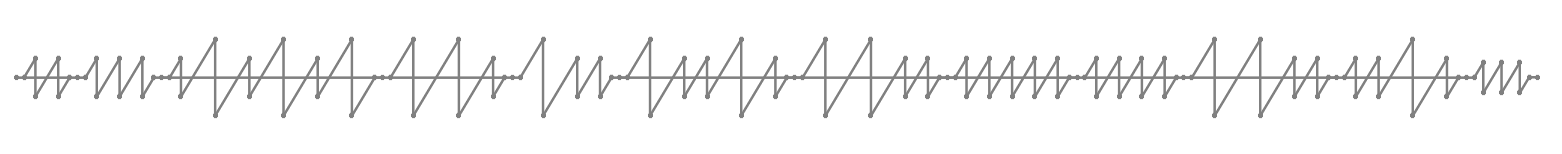}
     }
     \hfill
    \subfloat[\label{fig:hack3d2020_q1}]{%
       \includegraphics[width=0.6\columnwidth]{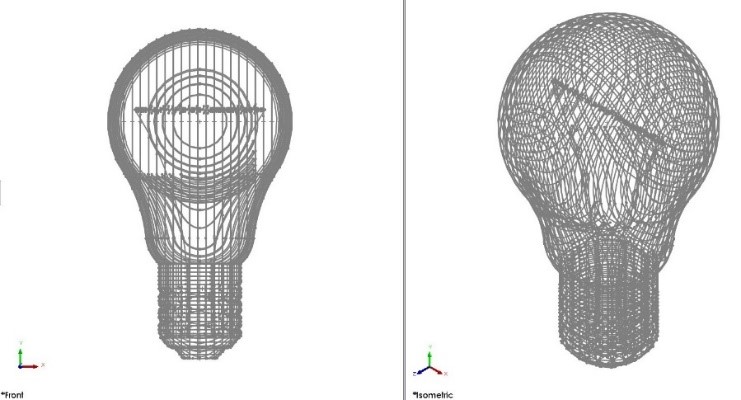}
     }
     \hfill
      \subfloat[\label{fig:hack3d2020_f1}]{%
       \includegraphics[width=0.5\columnwidth]{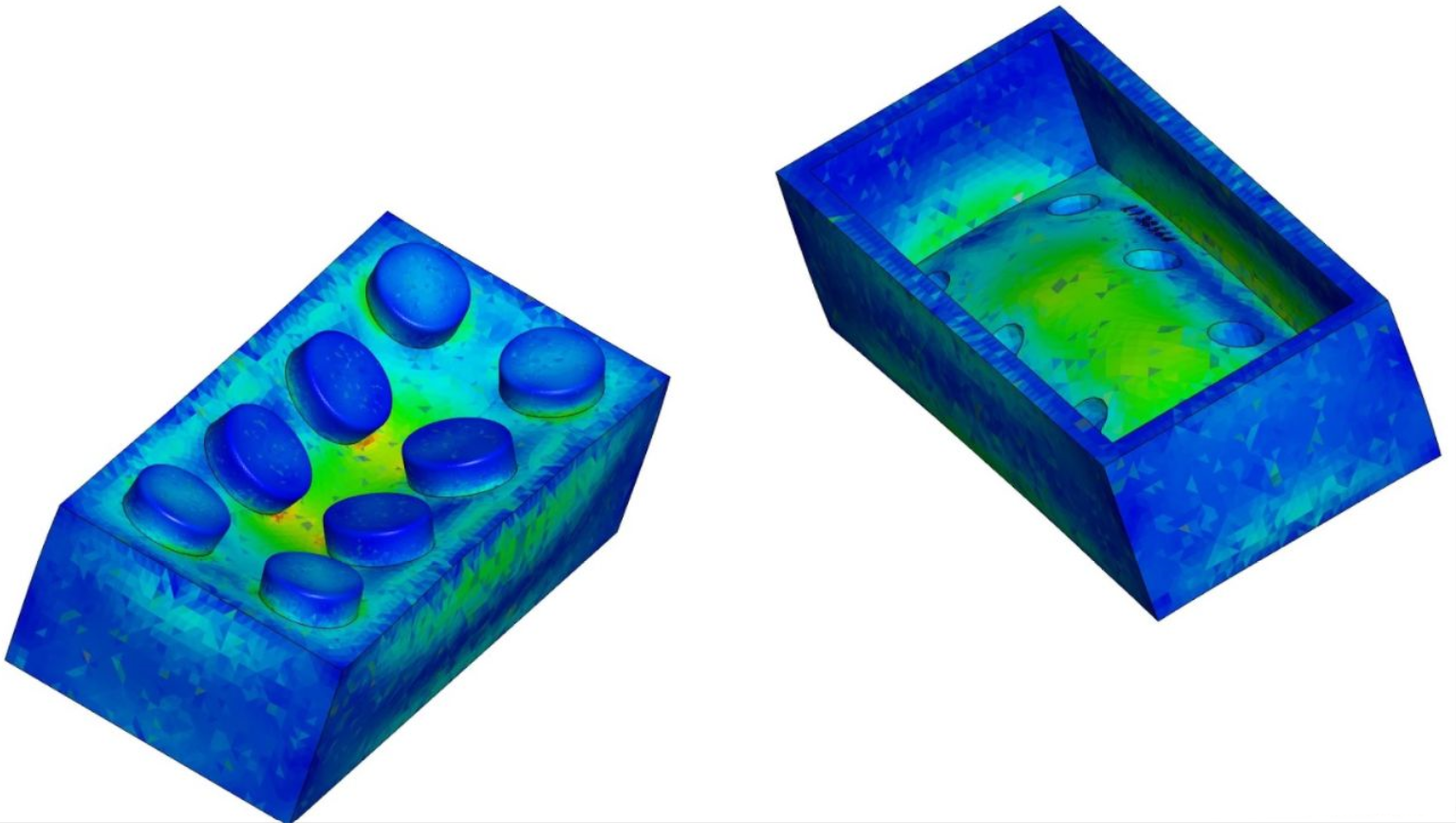}
     }
    \hfill
     \subfloat[\label{fig:hack3d2020_f2}]{%
       \includegraphics[width=0.7\columnwidth]{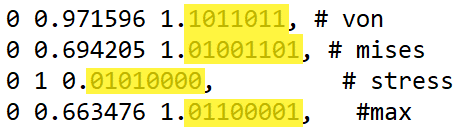}
     }
     \caption{In HACK3D Challenge 1, the participants were asked to reconstruct the 3D model (a) given a set of XYZ coordinates that describes the design. The coordinates can be visualized as a point cloud as in (b). In Challenge 2, the participants received STL files of designs A and B shown in (c). They had to find the correct orientation to slice and print that would eliminate surface and internal defects. If printed without rotation, design A will be separated in multiple segments as in (d) and design B will have internal slots as shown in (e). The correct orientations for printing and the supporting materials for designs A and B are shown in (f) and (g), respectively. In HACK3D Challenge 3, given (h) a partial (possibly damaged) G-code file, the attacker has to reconstruct the original G-code file that is cut off from a chess piece from among the (i) three candidates. If one views the 3D QR code embedded in the chess base from a random direction, it looks like a group of spheres like (j). However, viewed from the correct angle, it is a scannable QR code shown in (k). In HACK3D Challenge 4, the participants were given a female connector (in yellow) and a scaled-down version of the STL file of the female connector. They were challenged to use this information to reverse engineer the male part. Three reconstructed designs are shown in (l). In HACK3D Challenge 5, the participants had to recreate the lightbulb CAD based on the sketches in (m). In the center of the sketches, a unique code shown in (n) exists for participants to decode. The deformed results of the linear static simulation in HACK3D Challenge 6 is saved into a VRML file in (o). The maximum stress of the simulation is stored in a binary format and hidden in the color index section of the VRML file in (p).}
\end{figure*}

\noindent{\textbf{Challenge.}} The participants received a set of XYZ coordinates in the 3D space describing the shape of a part (see Figure~\ref{fig:hack3d_2018_target}). The red teams use this information to recreate a 3D model of the object. The XYZ coordinates can be visualized as a point cloud shown in Figure~\ref{fig:hack3d_2018_point_cloud}.

\vspace{2mm}
\noindent\textbf{Threat Scenario.} Challenge 1 exemplifies attacks that can be launched using the point cloud information of a design. These point cloud data can be generated from 3D scanners. This challenge shows the capability of computer visualization software to recover the CAD model from a point cloud file representing the part. Red-team participants take on the role of the adversary to develop their own reverse engineering method to recover the design files by using the point cloud information.

\vspace{2mm}
\noindent{\textbf{Attacks.}} One red team used Microsoft Excel to convert the coordinates into a point cloud and imported it into SolidWorks. They created a mesh using GeoMagic Add-in and obtained reference curves from the meshes. By combining the reference curves and the point cloud outline, the team  recreated the 3D model. 
Another red team used the FeatureScript tool in OnShape. The team developed scripts to extract information from the coordinates, draw poly lines, and delete unnecessary faces layer-by-layer. This way they recreated the 234 layers and assembled them to recreate the model.
A third red team of mechanical engineers used the Scanto3D tool in SolidWorks to reconstruct the 3D model.

\subsection{Challenge 2 (HACK3D 2018 Final Challenge)}

\noindent\textbf{Challenge.} The participants received stereolithography (STL) files of two designs (Figure~\ref{fig:hack3d_2018_final}) and were invited to identify the correct slicing and printing orientations that would remove all surface and internal defects. These two models were designed to have security features such that if they are not sliced in a specific direction using the required slicing parameters, the prints will have either internal or surface defects. 

\vspace{2mm}
\noindent\textbf{Threat Scenario.} This challenge mimics a real-world situation that STL files are somehow leaked to adversaries due to, e.g., disgruntled insiders or hacked file storage servers. However, the designers were able to embed defects in the design and want to prevent attackers from producing high-quality products just using the STL files. The intentional addition of the embedded defects is an example of an obfuscation countermeasure that was designed by the blue team. 

\vspace{2mm}
\noindent\textbf{Attacks.} Most teams sliced the models in different orientations to check whether the defects and nicks survive. Some participants created a table to enumerate all possible rotations. Since, we have x- y- and z-axis and since one can rotate by 360 degrees along these axes, there are $360^{3} \approx 4.7 \times 10^7$ combinations (considering a 1-degree rotation step). Although the brute force tries all these combinations, the red teams came up with strategies to narrow down the search space and get to the correct solution within the challenge time  of 7 hours.

First, the participants characterized design A as a complex, open prism with several holes. The flaws were introduced due to the presence of segmentation in each layer (Figure~\ref{fig:hack3d_segments}), which decreases structural integrity of the print. Their main goal was to find an orientation for design A such that each layer shows a continuous toolpath. Second, the surface area of the bottom layer is considered when the challenge participants were seeking the correct orientation. The bottom layer needs larger area and mass, so that it can improve the quality of printing; a larger bottom layer provides a superior adhesion to the base plate. After a few trials with the different orientations, some participants found the correct orientation for printing.

Design B is a solid box. However, inside the box, rectangular prisms were embedded with spade shaped flaws, giving the surface several nicks (Figure~\ref{fig:hack3d_2018_spade}). The participants discovered that the nicks remain if they do not turn the printing orientation, regardless of the choice of the bottom layer. By fine tuning the rotation angle, they eventually found the correct orientations for printing the Design B without internal defects. The correct printing orientations for Design A and B are shown in Figure~\ref{fig:hack3d_2018_finalA} and Figure~\ref{fig:hack3d_2018_finalB}, respectively.

\subsection{Challenge 3 (HACK3D 2019 Qualifying Challenge)}

\vspace{2mm}
\noindent\textbf{Challenge.} For this challenge, the red teams mimic attackers who steal a partially damaged G-code file, which only models the bottom part of a chess piece. Figure~\ref{fig:chess_base} shows the damaged G-code as viewed in a G-code viewer. The participants needed to solve two problems: (1) identify the correct piece among the three candidates pieces (Pawn, Bishop, Queen) that the partial design represents and (2) complete the design with the correct dimensions. The blue team provided an orthographic image of all candidates in Figure~\ref{fig:chess_pieces}, and a text file with true z-heights of each piece.

The blue team organizers embedded a non-trivial shortcut in the damaged G-code for solving the challenge. They placed the design file of the top half of the chess piece into a separate text file stored on the cloud, giving view-only access to those with a link. This link was embedded as a 3D quick response (QR) code in the design of the chess base given to the participants~\cite{chen2019embedding}.

\vspace{2mm}
\noindent\textbf{Threat Scenario.} This challenge is an example of a cybersecurity threat where the adversary is launching an attack on the design files to steal the design for counterfeiting.
Each participant of the HACK3D Challenge 3 took on the role of the red-team designer working on the G-code to  find the hidden information to recreate the complete file. The embedded QR code was used in the design to counter such direct sabotage attacks on the design files. 

\vspace{2mm}
\noindent\textbf{Attacks.} %
One of the teams exploited the discrepancy in the metadata in the G-code file. They noticed that the filament length in the original piece (4290.7 mm) was different from that shown by the G-code viewer (3198.14 mm). This offered insights into the cut-off design and they concluded that all pieces had a square cross-section. Next, using the provided z-heights and the extracted height of the base piece from the G-code, they determined where the chess piece was cut off. They cropped the tops off of each chess piece and used computer vision algorithms to measure the pixel dimensions and then scaled the dimensions using information from the G-code. They reconstructed the G-code for the top of all three pieces. Since they knew the height difference between the original and the damaged piece, they were able to deduce the Queen as the target piece, as it matched closest in height. The final result had an error of 1\%.

A second red team processed the image and created a profile of the edges of the chess pieces. This produced a 1\% error in the geometry as the image processing led to a pixelated line. Based on this chess piece edge information, the team created the shell of the Queen with the help of a few reference points in the G-code and filled the top and bottom layers with infill. 
The third team also used image processing methods. However, they took a different approach and produced a square prism at each point on the profile curve to recreate the pieces. They inferred that the chess piece was the Queen based on the filament length information in the damaged G-code.

Two of the teams recognized the QR code embedded in the chess base. This embedded QR code was segmented into small pieces and appear as a bundle of spheres shown in Figure~\ref{fig:spheres}. Only when viewed from a certain direction, the QR code could be seen as Figure~\ref{fig:qr_code}. One team extracted this QR code from the G-code and then extracted the design file of the chess piece stored on the cloud server.

\subsection{Challenge 4 (HACK3D 2019 Final Challenge)}

\noindent\textbf{Challenge.} targeted reverse engineering phase in the DM supply chain and file forensics. The red-team participants were given a physical print and a scaled-down version of the STL file of a female connector (yellow parts in Figure~\ref{hack3d_challenge4}). The challenge entailed construction of a male connector with the appropriate design and dimensions compatible with the female connector. Similar to Challenge 3, the design of the female connector had a 3D data matrix embedded within. This matrix, when viewed from the correct orientation, had the password of an online server whose internet protocol address and username were stored within the header of the STL file.%

\vspace{2mm}
\noindent\textbf{Threat Scenario} of this challenge is reverse engineering by an adversary who has stolen the design files. %
The adversary steals design file information for pirating and/or counterfeiting. %
Participants of this challenge played the role of the adversary and are tasked to create a male connector with only the female connector part information. Success of this challenge confirms that reverse engineering techniques may be used to obtain information about the missing/complementary components of a design from a stolen component.

\vspace{2mm}
\noindent\textbf{Attacks.} Under a tight time constraint of 6 hours, one red team was able to extract all the required information to access the design file stored in the server. The male part was designed with a snap fit and arms to prevent rotation (the left one in Figure~\ref{hack3d_challenge4}).
Another red team took a geometric approach and recreated a tight slide fit male part of the female part along with the scale factor. While they were able to get the data matrix, they could not recover the hidden message in the STL file. Hence they did not access the file stored on the server. Their final design is the right most one in Figure~\ref{hack3d_challenge4}. 
Finally, one team manipulated the STL file and isolated a single cross-section of triangles to create a profile of the female part. After fine-tuning the profile, they conducted multiple design and printing iterations of a snug fit to slide on the male connector (the middle one in Figure~\ref{hack3d_challenge4}). Such brute force approach was very time and material intensive as 3D printing of the part took more than half an hour each time. 

\subsection{Challenge 5 (HACK3D 2020 Qualifying Challenge)}

\noindent\textbf{Challenge.} Participants were given an Initial Graphics Exchange Specification (IGES) file of a lightbulb design that had all its solid body geometry removed and replaced with sketches. This challenge was to reverse engineer the lightbulb CAD model from these sketches. The sketches provided an overview of how the part should look like, but it is hard to determine the dimensions precisely. Further, a hidden Morse code sketch was embedded into the center of the bulb to allow the participants to obtain the original 2D drawing of the lightbulb if it was was decoded correctly. Figure~\ref{fig:hack3d2020_q2} shows this hidden code, where the short vertical lines represent dots and the long vertical lines represent dashes in the Morse code. The IGES file is shown in Figure~\ref{fig:hack3d2020_q1}, it only has 3D sketches that represent the silhouette of the lightbulb design created from planar section cuts of the actual model.

\vspace{2mm}
\noindent\textbf{Threat Scenario.} The main purpose of this challenge was to simulate the scenario in which an attacker launches IP theft attack on the engineering design files. The blue team saved the design file with a unique storage method and the red teams had to recover these files. In this challenge, the initial feedback from some of the teams was that the IGES file was opening as an empty, broken file. The import options of the software had to be modified to be able to view the IGES file correctly. One aspect of this challenge is to help introduce the idea of playing with the storage method to better protect design information. %

\vspace{2mm}
\noindent\textbf{Attacks.} Most teams approached the challenge by approximating the measurements of the features of the lightbulb and recreating the part based on the measured dimensions. One team decided to view the IGES file in a text editor and was able to recover the information about the original design file. The team proceeded to extract the center plane sketches to determine the radius of the glass bulb. They recreated the remaining features by analyzing the small sections of the sketches and recovered the dimensions of these features. 

The challenge file had a lot of sketches that had to be properly filtered and analyzed. Otherwise, it can lead to wrong inferences about the features in the design. The multiple cross-sectional planar sketches misled one red team in designing asymmetric support rods. They used this incorrect assumption in their design. The same team successfully determined that the filament had no cross-section thickness sketch but it did not evoke suspicion. They guessed the cross section was circular but could not conclude that the filament was a coded message.  

\subsection{Challenge 6 (HACK3D 2020 Final Challenge)}

\noindent\textbf{Challenge.} A linear-static FEA simulation was conducted on the challenge model and the results were exported to a Virtual Reality Modeling Language (VRML) file as shown in Figure~\ref{fig:hack3d2020_f1}. The VRML file shows a graphics body representation of the deformed part, containing the stress result colors and facets from the mesh of the simulation. Based on the given data, the red teams were tasked to determine the maximum value of the FEA and to add back the missing interior connectors.

In the interior of the part, the blue team inscribed a code as an insignificant part number. This number references a line in the VRML file when viewed in a text editor. The line corresponds to the color index in the file and this is where the red, green, and blue (RGB) values of each facets are stored. The value and units of measurement of the maximum stress of the simulation were converted into binary value and stored in the VRML file. The binary values replaced the digits that came after the decimal for the green intensity value of the color index as shown in Figure~\ref{fig:hack3d2020_f2}. 

\vspace{2mm}

\noindent\textbf{Threat Scenario.} The attack depicted in this challenge is malicious reduction in structural integrity of the part. An adversary can access the design files and inject defects into the model that is then carried out into the manufacturing stage. If the defect is not detected, the production parts will have the defect in them and the parts can malfunction during operation. The blue team used this challenge to assess the effectiveness of storing simulation data within design files. The set of original simulation results hidden within the design file can be retrieved to verify the integrity of the part.

\vspace{2mm}
\noindent\textbf{Attacks.} Only one team was able to follow the clues in the VRML file, which led them to the line containing the binary values. They successfully determined the value of the maximum stress in the simulation of the part. Two teams followed a different approach of conducting a new simulation study to obtain the maximum stress value. The teams had to generate a new CAD file because they could not use the deformed model in a simulation analysis. It was difficult to perform the simulation study because there was a lot of missing information about the simulation input data. The teams had to guess the applied load, material, and correct fixtures for the part. 

A team of computer science students realized that the file stores the RGB value of each facet in the file and the maximum stress values corresponds to the facet that are red. They located two color indices that had the red color in the VRML file. They tried to relate them to other information in hopes to obtain the maximum stress. While the team failed in uncovering any further helpful information within the time constraints, this was a very creative approach.

\subsection{Statistics}

Twenty-four red teams registered to compete in the Challenge 3 round of the 2019 HACK3D. Each team had 2-4 students, who were pursuing a degree in either mechanical engineering, computer science or computer engineering. Five teams advanced to the final round. In the 2020 HACK3D Qualifying Challenge 5, 43 teams from around the globe registered to compete. We expect this trend in increase in participation from around the world is expected for future HACK3D competitions. As more teams participate in these challenge, the blue team organizers can compile innovative attacks and benchmarks to evaluate the security methods and uncover new attack vectors through crowd-sourcing.  

\subsection{Lessons Learned}

\begin{table*}[]
\centering
\caption{Summary of Attack Methods Proposed by HACK3D Teams.}\label{tab:hack3d}
\resizebox{\textwidth}{!}{
\begin{tabular}{|c|c|c|c|c|}
\hline
            & \textbf{Threat Taxonomy (Goal/Method/Target)}                 & \textbf{Countermeasure} & \textbf{Information Exploited}                                                                                      & \textbf{Red team Skills}                                                                                   \\ \hline
Challenge 1 & Counterfeit/Reverse Engineer/CAD (point cloud)    & Obfuscation    & Geometric Information                                                                                      & Reverse Engineering                                                                      \\ \hline
Challenge 2 & Sabotage/Tamper Data/STL                             & Obfuscation    & \begin{tabular}[c]{@{}c@{}}Geometric Information\\    Printing Parameters\end{tabular}                     & \begin{tabular}[c]{@{}c@{}}CAD\\    3D Printing\end{tabular}                             \\ \hline
Challenge 3 & Sabotage/Tamper Data/G-code                          & Authentication & \begin{tabular}[c]{@{}c@{}}Metadata of G-code\\    Geometric Information\\    Hidden Code\end{tabular}     & \begin{tabular}[c]{@{}c@{}}Image Processing\\    File Manipulation\\    CAD\end{tabular} \\ \hline
Challenge 4 & Counterfeit/Reverse Engineer/CAD (physical print) & Fingerprinting & \begin{tabular}[c]{@{}c@{}}Physical Measurements\\    Hidden Code\end{tabular}                             & \begin{tabular}[c]{@{}c@{}}CAD\\    File Manipulation\\    3D Printing\end{tabular}      \\ \hline
Challenge 5 & Counterfeit/Reverse Engineer/CAD (IGES)           & Obfuscation    & \begin{tabular}[c]{@{}c@{}}Geometric Information\\    Hidden Code\end{tabular}                             & CAD                                                                                      \\ \hline
Challenge 6 & Sabotage /Tamper Data/STL (Simulation)               & Authentication & \begin{tabular}[c]{@{}c@{}}Geometric Information\\    Simulation Information\\    Hidden Code\end{tabular} & \begin{tabular}[c]{@{}c@{}}CAD\\    File Manipulation\\    FEA\end{tabular}              \\ \hline
\end{tabular}
}
\end{table*}

Table~\ref{tab:hack3d} maps the challenges to the threat taxonomy and summarizes the skill sets used by participants. By analyzing the performance of the blue and read teams, the following lessons were learned

\begin{enumerate}
    \item \textbf{More information can be extracted from leaked files than what is anticipated.} For example, in HACK3D Challenge 3, one team looked into the metadata of the corrupted G-code and extracted valuable information. In Challenge 5, one team determined information about the software and designer that created the original CAD file. 
    \item \textbf{Prior 3D printing and CAD experiences can be advantageous in reverse engineering attacks.} In HACK3D Challenge 2, past experiences in 3D printing helped in the attacks. Theoretically, there are around 50 million possible angle combinations, but an attacker with rich printing experiences can quickly rule out many combinations. In Challenge 4, the red team's individual design experiences led to different male connector CAD models.
    \item \textbf{Attackers need not necessarily be experts in DM %
    to launch a successful attack.} In HACK3D Challenge 1, commercial CAD and add-in tools gave participants without any experiences an advantage. These software can aid attackers in reverse engineering situations.
    \item \textbf{Multi-disciplinary knowledge and skills are useful both from an attacker and a defenders perspective.} Although one does not require a deep understanding of cybersecurity to launch an attack on DM systems and supply chains, more sophisticated and novel attacks can be developed if one can combine knowledge and expertise from different disciplines like computer science, electrical engineering, mechanical engineering, and material science. This is also why HACK3D challenges strongly encourage participants who had different technical backgrounds to join forces with each other and form cross-disciplinary teams. The skills employed by the teams in the attacks are listed in Table~\ref{tab:hack3d}.
\item \textbf{Attacks are not created equal.} Each HACK3D challenge asked the participants to achieve the same attack goal, so all the successful attacks in each challenge completed the same purpose. However, since the participants tackled the same problem from various angles, the developed attacks require different countermeasures. For example, reverse engineering can be thwarted by design obfuscation, and secret file leakage requires stronger access control and authentication in the information technology (IT) system.

    \item \textbf{Attacks can originate in any stage in the DM supply chain.} The HACK3D challenges show that attackers can launch attacks at any stage in the DM supply chain, including STL files, IGES files, G-code files, and the physical prints.  DM security researchers should design and deploy security measures to protect the DM supply chain end-to-end. 
    \item {\bf The taxonomy outlines the numerous defenses and attack pathways.} In the 3 years of HACK3D, we explored a small set of pathways through the taxonomy. We exhort the emerging manufacturing cybersecurity community to study the unexplored threat taxonomy pathways. NYU Center for Cybersecurity will continue to do so in future  HACK3D challenges. 
    \item {\bf There is a huge space for the attackers to explore and exploit.} The HACK3D challenges follow the philosophy in~\cite{forbes2020investigating}, and were designed to unleash the imagination of attackers. Only the attack targets were defined by each HACK3D challenge, and the participants are free to find their own way to accomplish the goal. Participants often surprised the challenge designers with their creative attacks. For example, the information leakage from the metadata in a G-code file was unexpected.
\end{enumerate}

\section{Conclusions} \label{s:conclusion} 
Securing the DM cyber-physical system is a challenging task. 
We are conducting an annual crowdsourcing red-team-blue-team event to assess the strength of the DM security methods yielding novel attacks on them. %
While it is in its formative years, HACK3D showed that red teams with a range of skills --with minimal knowledge in DM and cyber security to expert interdisciplinary knowledge-- can develop innovative attacks in defeating the embedded security. The defenses and the attacks can be used to benchmark both future defenses and attacks for the DM community. The approaches documented by HACK3D offer insights into the next generation of DM security methods and their application in the DM manufacturing processes. Consistently, we noticed that the red team participants obtained more information from the artifacts than we anticipated and this informed effective attacks. Despite a stringent timeline for solving the HACK3D challenges, the red teams made significant advancements and many of them solved the challenges. Clearly, a multi-disciplinary training is important for the emerging DM workforce to develop DM-unique security methods for DM CPS. Otherwise they may not anticipate many of the impending attack vectors. HACK3D will continue in Fall 2021 and we project over 60 teams (and 180 students) to participate. Future HACK3D challenges will investigate unexplored pathways through the DM threat taxonomy. All benchmarks~\footnote{https://github.com/NYUCSAW/HACK3D} from these HACK3D challenges can be used by the DM community to improve defenses and train the next generation of DM workforce.
From our experience with organizing CSAW Capture the Flag and the embedded security challenges \cite{karri2010trustworthy}, we are optimistic that HACK3D attacks and defenses will become the basis for an open, community-accessible benchmark resource that the DM community can use, add onto, and improve DM cybersecurity.

\section*{Acknowledgement}

HACK3D events are supported by the National Science Foundation Secure and Trustworthy Computing grant DGE-1931724. Chenglu was supported by NYU CCS and NYU CUSP. The views presented in the paper are those of authors, not of the funding agency. We thank NYU-Tandon Makerspace for supporting HACK3D. We thank all red teams and the judges for  participating in HACK3D.%

\bibliographystyle{IEEEtran}
\bibliography{additive}

\begin{IEEEbiographynophoto}{Michael Linares}
is an undergraduate Mechanical Engineer at NYU Tandon School of Engineering. In the Composite Materials and Mechanics Lab, he has researched novel methods of protecting the AM industry. %
Contact him at michael.linares@nyu.edu.
\end{IEEEbiographynophoto}

\begin{IEEEbiographynophoto}{Nishant Aswani}
is a Computer Engineering undergraduate at New York University Abu Dhabi. %
His areas of experience and interest lie in various 3D modelling problem, machine learning and vision, and automation. Contact him at nsa325@nyu.edu.
\end{IEEEbiographynophoto}

\begin{IEEEbiographynophoto}{Gary Mac}
is a doctoral student of Mechanical and Aerospace Engineering at New York University. He graduated with a B.S./M.S. degree in Mechanical Engineering at New York University. His research is focused on digital manufacturing security and developing secure and trustworthy computer aided design files to counter threats in the 3D printing process chain. Contact him at gm1247@nyu.edu.%
\end{IEEEbiographynophoto}

\begin{IEEEbiographynophoto}{Chenglu Jin}
is a tenure-track researcher in the Computer Security Group at CWI Amsterdam. He holds a Ph.D. degree from the University of Connecticut. His research interest is cyber-physical system security, hardware security, and applied cryptography. Contact him at chenglu.jin@cwi.nl.  
\end{IEEEbiographynophoto}

\begin{IEEEbiographynophoto}{Fei Chen} is currently working for Apple in China.
She received her Ph.D. in Mechanical Engineering from NYU Tandon School of Engineering in 2019. Her area of expertise is additive manufacturing security, including secure CAD model files and embedded codes for product authentication. Contact her at fionachane@gmail.com.
\end{IEEEbiographynophoto}

\begin{IEEEbiographynophoto}{Nikhil Gupta}
is a Professor of Mechanical and Aerospace Engineering at New York University. He is also affiliated with NYU Center for Cybersecurity. He received his Ph.D. from Louisiana State University in Mechanical Engineering and M.S. from Indian Institute of Science in Materials Science. His research is focused on cybersecurity and machine learning in manufacturing. %
He is a senior member of IEEE. Contact him at ngupta@nyu.edu.
\end{IEEEbiographynophoto}

\begin{IEEEbiographynophoto}{Ramesh Karri}
is  a Professor of Electrical and Computer Engineering with New York University (NYU), Brooklyn, NY, USA. He co-directs and co-founded the NYU Center for Cyber Security. He holds a Ph.D. degree from the University of California, San Diego. %
His research interests include hardware cybersecurity include trustworthy ICs; processors and cyberphysical systems; security-aware computer-aided design, test, verification, validation, and reliability; nano meets security; hardware security competitions, benchmarks and metrics; biochip security; and additive manufacturing security. He is a Fellow of the IEEE. Contact him at rkarri@nyu.edu.
\end{IEEEbiographynophoto}

\end{document}